\def\article{0}
\def\lncs{1}
\def\sigalternate{2}

\def\ieee{4}
\def\style{\lncs}

\ifnum\style=\article
 \documentclass[11pt,letter]{article}
 \usepackage{fullpage}
\usepackage{fixltx2e}
\usepackage[activate]{pdfcprot}
\usepackage{amsthm,amsfonts}
\usepackage{wrapfig}
\usepackage{xspace}
\usepackage{amsmath} 
\theoremstyle{plain}
\usepackage{algorithmic}
\newtheorem{theorem}{Theorem}
\newtheorem{lemma}[theorem]{Lemma}
\newtheorem{claim}[theorem]{Claim}
\newtheorem{corollary}[theorem]{Corollary}
\newtheorem{definition}{Definition}

\theoremstyle{definition}
{\bfseries}{\itshape}
\newtheorem{remark}{Remark}
\newtheorem{example}[theorem]{Example}

\fi
\ifnum\style=\lncs
 \let\accentvec\vec 
 \documentclass{llncs}
 \let\vec\accentvec 
 \usepackage{amsmath,amsfonts}
 \usepackage{wrapfig}
\fi
\ifnum\style=\sigalternate
\usepackage{algorithmic}
\newcounter{theorem}
\newcounter{definition}
\newcounter{claim}
\newcounter{remark}
\newtheorem{lemma}{Lemma}
\newtheorem{corollary}{Corollary}

\newdef{mydefinition}{Definition}
\newdef{construction}{Construction}
\newdef{example}{Example}

\newenvironment{definition}[1]{
  \refstepcounter{definition}
  \medskip
  \indent
  \textit{Definition \thedefinition\ (\textit{#1}).}
}{}

\newenvironment{remark}{
  \refstepcounter{remark}
  \medskip
  \indent
  \textit{Remark \theremark.}
}{}

\newenvironment{claim}{
  \refstepcounter{claim}
  \medskip
  \indent
  \textit{Claim \theclaim.}
}{}

\newenvironment{theorem}[1]{
  \refstepcounter{theorem}
  \medskip
  \indent
  \textsc{Theorem \thetheorem}\ (\textit{#1}).
}{}

\fi
\ifnum\style=\ieee
\documentclass[conference]{IEEEtran}

\newtheorem{definition}{Definition}

\usepackage{amssymb}
\usepackage{amsmath}
\fi

\usepackage{ifthen}
\usepackage[ruled,vlined]{algorithm2e}
 \usepackage[pdftex,colorlinks,backref]{hyperref}
 \hypersetup{colorlinks=false, menucolor=blue, urlcolor=blue}
 \renewcommand*{\backref}[1]{}
\usepackage{graphicx,xcolor}
\usepackage{tabularx}
\usepackage{booktabs}
\usepackage[utf8]{inputenc}
\usepackage{paralist}

\usepackage{centernot}

\usepackage{multicol}
\usepackage{algorithmic}
\usepackage{url}
\usepackage{graphicx,color}
\usepackage[american]{babel}
\usepackage{listings}
\usepackage{paralist}
\graphicspath{ {./pic/} }

\newboolean{ANONYMOUS}
\setboolean{ANONYMOUS}{FALSE}
\newcommand{\bin}{\{0,1\}}
\newcommand\bit{\{0,1\}}

\newcommand{\secparam}{{\ensuremath{\lambda}}}

\newcommand{\exec}{\ensuremath{\leftarrow}}

\newcommand{\olrk}[1]{%
   \ifx\nursymbol#1\else\!\!\mskip4.5mu plus 0.5mu\left(#1\right)\fi}
\newcommand{\elrk}[1]{%
   \ifx\nursymbol#1\else%
        \!\!\mskip4.5mu plus0.5mu\left[\mskip2.5mu plus0.5mu #1\right]\fi}

\newcommand{\mypar}[1]{\parskip 5pt
\noindent{\underline{\textsc{#1:}}}}

\newcommand{\newsequenceofgames}[1]{
  \newcounter{#1}
  \setcounter{#1}{-1}

  \ifx \GameID \undefined
    \newcommand{\GameID}{#1}
  \else
    \renewcommand{\GameID}{#1} 		
  \fi

  \ifx \PrevLabel \undefined
    \newcommand{\PrevLabel}{\GameID.NULL}
  \else
    \renewcommand{\PrevLabel}{\GameID.NULL}
  \fi

  \ifx \ThisLabel \undefined
    \newcommand{\ThisLabel}{\GameID.NULL}
  \else
    \renewcommand{\ThisLabel}{\GameID.NULL}
  \fi
}

\newcommand{\nextgame}[1]{
  \let\PrevLabel\ThisLabel
  \renewcommand{\ThisLabel}{\GameID.#1}
  \refstepcounter{\GameID}\label{\GameID.#1}
  \paragraph{Game~\arabic{\GameID}.}
}

\definecolor{lightgray}{gray}{0.9}
\definecolor{gray}{rgb}{0.5, 0.5, 0.5}
\definecolor{personcolor}{rgb}{0.9, 0.7, 0.7}
\usepackage{xcolor,colortbl}

\makeatletter\newenvironment{graybox}{%
\begin{lrbox}{\@tempboxa}\begin{minipage}{\columnwidth}}{\end{minipage}\end{lrbox}%
\colorbox{lightgray}{\usebox{\@tempboxa}}
}\makeatother



\newenvironment{attention}{%
\begin{tabular}{m{4mm}|m{2mm}m{115mm}}
\textbf{\rotatebox{90}{\mbox{Note}}} & &
\begin{minipage}[t]{\linewidth}
}{
	\end{minipage}
	\end{tabular}
}

\newcommand{\encKey}{\textsf{key}}
\newcommand{\did}{\textsf{did}}
\newcommand{\sid}{\textsf{sid}}

\newcommand{\cid}{\textsf{tid}}
\newcommand{\aid}{\textsf{uid}}

\newcommand{\layout}{\textsf{layout}}
\newcommand{\signature}{\textsf{signature}}
\newcommand{\encCh}{\textsf{encrypted challenge}}
\newcommand{\timestamp}{\textsf{timestamp}}

\newcommand{\decode}{\ensuremath{\textsc{Decode}}}
\newcommand{\encode}{\ensuremath{\textsc{Encode}}}
\newcommand{\token}{\ensuremath{\textsc{Token}}}
\newcommand{\veriDoc}{VeriDoc}
\newcommand{\sevi}{Ubic}
\newcommand{\pke}{\ensuremath{\Pi_\mathsf{pke}}}
\newcommand{\pkegen}{\ensuremath{\mathsf{Gen}}}
\newcommand{\encrypt}{\ensuremath{\mathsf{Enc}}}
\newcommand{\decrypt}{\ensuremath{\mathsf{Dec}}}
\newcommand{\ek}{\ensuremath{\textit{ek}}}
\newcommand{\dk}{\ensuremath{\textit{dk}}}
\newcommand{\ocr}{\mathsf{OCR}}
\newcommand{\privke}{\ensuremath{\Pi_{\mathsf{priv}}}}
\newcommand{\privkegen}{\ensuremath{\mathcal{G}}}
\newcommand{\privencrypt}{\ensuremath{\mathcal{E}}}
\newcommand{\privdecrypt}{\ensuremath{\mathcal{D}}}

\newcommand{\emx}[1]{\ensuremath{#1}\xspace}
\newcommand{\isProf}{\emx{\mathsf{mayAccProf}}}
\newcommand{\isEmployee}{\emx{\mathsf{mayAccEmp}}}
\newcommand{\isStud}{\emx{\mathsf{mayAccStud}}}
\newcommand{\Employee}{\emx{\mathit{Emp}}}
\newcommand{\file}{\emx{\mathit{file}}}
\newcommand{\Prof}{\ensuremath{\mathit{Prof}}\xspace}
\newcommand{\Stud}{\ensuremath{\mathit{Stud}}\xspace}

\newcommand{\mc}[1]{\ensuremath{\mathcal{#1}}\xspace}
\newcommand{\mbb}[1]{\ensuremath{\mathbb{#1}}\xspace}

\newcommand{\generator}[1]{\ensuremath{\langle #1 \rangle}\xspace}
\newcommand{\set}[1]{\ensuremath{\{#1\}}\xspace}

\newcommand{\PE}{\ensuremath{\Pi_\mathsf{PE}}\xspace}
\newcommand{\PGen}{\ensuremath{\mathsf{PrGen}}\xspace}
\newcommand{\PKeyGen}{\ensuremath{\mathsf{PrKGen}}\xspace}

\newcommand{\PEnc}{\ensuremath{\mathsf{PrEnc}}\xspace}
\newcommand{\PDec}{\ensuremath{\mathsf{PrDec}}\xspace}
\newcommand{\mpk}{\ensuremath{\mathit{mpk}}\xspace}
\newcommand{\msk}{\ensuremath{\mathit{psk}}\xspace}
\newcommand{\fmsk}[1]{\ensuremath{{\mathit{sk}_{#1}}}\xspace}

\newcommand{\POGen}{\ensuremath{\mathsf{PoGen}}\xspace}
\newcommand{\POKeyGen}{\ensuremath{\mathsf{PoKGen}}\xspace}
\newcommand{\POEnc}{\ensuremath{\mathsf{PoEnc}}\xspace}
\newcommand{\PODec}{\ensuremath{\mathsf{PoDec}}\xspace}

\newcommand{\mopk}{\ensuremath{\mathit{opk}}\xspace}
\newcommand{\mosk}{\ensuremath{\mathit{osk}}\xspace}
\newcommand{\fmosk}[1]{\ensuremath{{\mathit{osk}_{#1}}}\xspace}

\newcommand{\vk}{\ensuremath{\mathsf{vk}}}
\newcommand{\sk}{\ensuremath{\mathsf{sk}}}
\newcommand{\skeygen}{\ensuremath{\mathsf{Kg}_{\mathsf{Sig}}}}
\newcommand{\sverify}{\ensuremath{\mathsf{Vf}}}
\newcommand{\ssign}{\ensuremath{\mathsf{Sig}}}
\newcommand{\sigscheme}{\ensuremath{\mathsf{DS}}}

\title{
\sevi: Bridging the Gap Between Digital Cryptography and the Physical World
}

\ifthenelse{\boolean{ANONYMOUS}}{
\author{Authors removed for blind review.}
}{
\author{Mark Simkin\inst{1}  \and Dominique Schr{\"o}der\inst{1} \and Andreas Bulling\inst{2} \and Mario Fritz\inst{2} }
\institute{Saarland University \\Saarbr\"ucken, Germany \and Max Planck Institute for Informatics \\Saarbr\"ucken, Germany}
\date{}
}

\begin{document}
%
%
%
%
%
%
%
%
%
%
\maketitle


\begin{abstract}
Advances in computing technology increasingly blur the boundary between the digital domain and the physical world. 
Although the research community has developed a large number of cryptographic primitives and has demonstrated their usability 
in all-digital communication, many of them have not yet made their way into the real world due to usability aspects. 
We aim to make another step towards a tighter integration of digital cryptography into real world interactions. 
We describe \sevi, a framework that allows users to bridge the gap between digital cryptography and the physical world. \sevi~relies on head-mounted displays, like Google Glass, resource-friendly computer vision techniques as well as mathematically sound cryptographic primitives to provide users with better security and privacy guarantees. The framework covers key cryptographic primitives, such as secure identification, document verification using a novel secure physical document format, as well as content hiding. To make a contribution of practical value, we focused on making Ubic as simple, easily deployable, and user friendly as possible.

\end{abstract}

\section{Introduction}\label{sec:intro}

Over the past years, the research community has developed a large number of cryptographic primitives and has shown their utility in all-digital communication. Primitives like signatures, encryption schemes, and authentication protocols have become commonplace nowadays and provide mathematically proven security and privacy guarantees. In the physical world, however, we largely refrain from using these primitives due to usability reasons. Instead, we rely on their physical counterparts, such as hand-written signatures, which do not provide the same level of security and privacy. Consider the following examples:

\paragraph{Authentication.}
In practice, most systems, such as ATMs or entrance doors, rely on the two-factor authentication paradigm, where a user, who wants to authenticate himself, needs to provide a possession and a knowledge factor. 
At an ATM, for instance, the user needs to enter his bank card and a PIN in order to gain access to his bank account.
Practice has shown that this type of authentication is vulnerable to various attacks~\cite{infectedUSB,skimming,targetBreach}, such as skimming, where the attacker mounts a little camera that films the PIN pad and a fake card reader on top of the actual card reader that copies the card's content.
Here, the fact that users authenticate with fixed credentials is exploited to mount large scale attacks by attacking the ATMs rather than specific users.

\paragraph{Hand-written signatures.}
Physical documents with hand-written signatures are the most common form of making an agreement between two or more parties legally binding.
In contrast to digital signatures, hand-written signatures do not provide any mathematically founded unforgeablility guarantees.
Furthermore, there is no well-defined process of verifying a hand-written signature.
This would require external professional help, which is expensive, time consuming, and therefore not practical.

\paragraph{Data privacy.}
Todays workplace is often not bound to specific offices or buildings any more. 
Mobile computing devices allow employees to work from hotels, trains, airports, and other public places.
Even inside office buildings, novel working practices such as 'hot-desking'~\cite{hotDesking} and 'bring your own device'~\cite{byod} are employed more and more to increase the employee's satisfaction, productivity, and mobility.
However, these new working practices also introduce new privacy threats. 
In a mobile working environment, potentially sensitive data might be leaked to unauthorized individuals, who can see the screen of the device the employee is working on. 
A recent survey~\cite{visualSecurity} of IT professionals shows that this form of information theft, known as \emph{shoulder surfing}, constantly gains importance. 
85\% of those surveyed admitted that they have at least once seen sensitive information that they were not supposed to see on somebody else's screen in a public place.
80\% admitted that it might be possible that they have leaked sensitive information at a public place.

\paragraph{}
In this work, we present \sevi, a framework and prototype implementation of a system that allows users to bridge the gap between digital cryptography and the physical world for a wide range of real world applications. \sevi~relies on \emph{head-mounted displays} (HMDs), like Google Glass~\footnote{\url{https://www.google.com/glass/}}, resource-friendly computer vision techniques as well as mathematically sound cryptographic primitives to provide users with better security and privacy guarantees in all of the scenarios described above in a user-friendly way.
Google Glass consists of a little screen mounted in front of the user's eye and a front-facing camera that films the users view.
It supports the user in an unobtrusive fashion by superimposing information on top of the users view when needed.

\subsection{Contributions}\label{sec:contribution}

To make a contribution of practical importance, in this work we focus on providing a resource-friendly, easy-to-use system, that can be seamlessly integrated into the current infrastructures. \sevi~offers the following key functionalities:

\paragraph{Authentication.}
We use a HMD in combination with challenge-response protocols to allow users to authenticate themselves in front of a device, such as an ATM or a locked entrance door. In contrast to current solutions, the PIN is not fixed but generated randomly each time.
Neither does an attacker gain any information from observing an authentication process, nor does he gain any from compromising the ATM or the bank, since they can only generate challenges, but not solve them. 
Copying the card does not help the attacker, since it does not contain any secret information, but merely a public identifier.

\paragraph{Content Verification.}
We enable the generation and verification of physical contracts with mathematically proven unforgeability guarantees.
For this purpose, we propose a new document format, \veriDoc, that allows for robust document tracking and optical character recognition, and contains a digital signature of its content. Using the HMD, a user can conveniently and reliably verify the validity of the document's content. 

\paragraph{Two-Step Verification.}
Based on the signature functionality described above, we introduce \emph{two-step verification} of content. During an online banking session, for instance, a user might request his current account balance. This balance is then returned along with a signature thereof. Using the HMD, we can verify the signature, and therefore verify the returned account balance. In this scenario, an attacker would need to corrupt the machine that is used for the banking session and the HMD at the same time in order to successfully convince the user of a false statement.

\paragraph{Content Hiding.}
We provide a solution for ensuring privacy in the mobile workplace setting.
Rather than printing documents in plain, we print them in an encrypted format. 
Using the HMD, the user is able to decrypt the part of the encrypted document that he is currently looking at. 
An unauthorized individual is not able to read or decrypt the document without the corresponding secret key.
Companies commonly allow employees with certain security clearances to read certain documents. 
We use predicate encryption to encrypt documents in such a way that only employees with the requested security clearances can read them.

\subsection{Smartphones vs. Head-Mounted Displays}
It might seem that all of the above scenarios could also be realized with a smartphone. 
This is not the case. In the content hiding scenario, we rely on the fact that the decrypted information is displayed to the user right in front of his eye.  A smartphone is still vulnerable to shoulder-surfing and would therefore not provide any additional privacy guarantees.
Realizing the authentication scenario with a smartphone is also problematic because a loss of possession is hard to detect and therefore an attacker might gain access to all secret keys as soon as he obtains the phone. 
Requiring the user to unlock the phone before each authentication process does not solve the problem, since the attacker might simply observe the secret that is used to unlock the phone.

\sevi\ overcomes this problem by using the so-called \emph{on-head detection} feature of the Google Glass device\footnote{\url{https://support.google.com/glass/answer/3079857}}.
The device is notified whenever it is taken off and at this point, \sevi\ removes all keys from the memory and only stores them in an encrypted format on internal memory.  HMDs are considered to be companions that are worn the whole day and only used when needed.
When a user puts on the device in a safe environment, he has to unlock it once through classical password entry.
Future versions might be equipped with an eye tracker, which would allow gaze-based password entry~\cite{GazePassShoulderSurfing}.


\section{The \sevi~Framework}
\label{sec:technical_details}

The key aim of \sevi~is to provide a contribution of practical importance that bridges the gap between digital cryptography and real world applications. We put emphasis on making our solutions as simple as possible and only use well researched and established cryptographic primitives in combination with resource friendly computer vision techniques to allow for easy deployment and seamless integration into existing infrastructures. 


\begin{figure*}[t]
\centering
  \includegraphics[width=\linewidth]{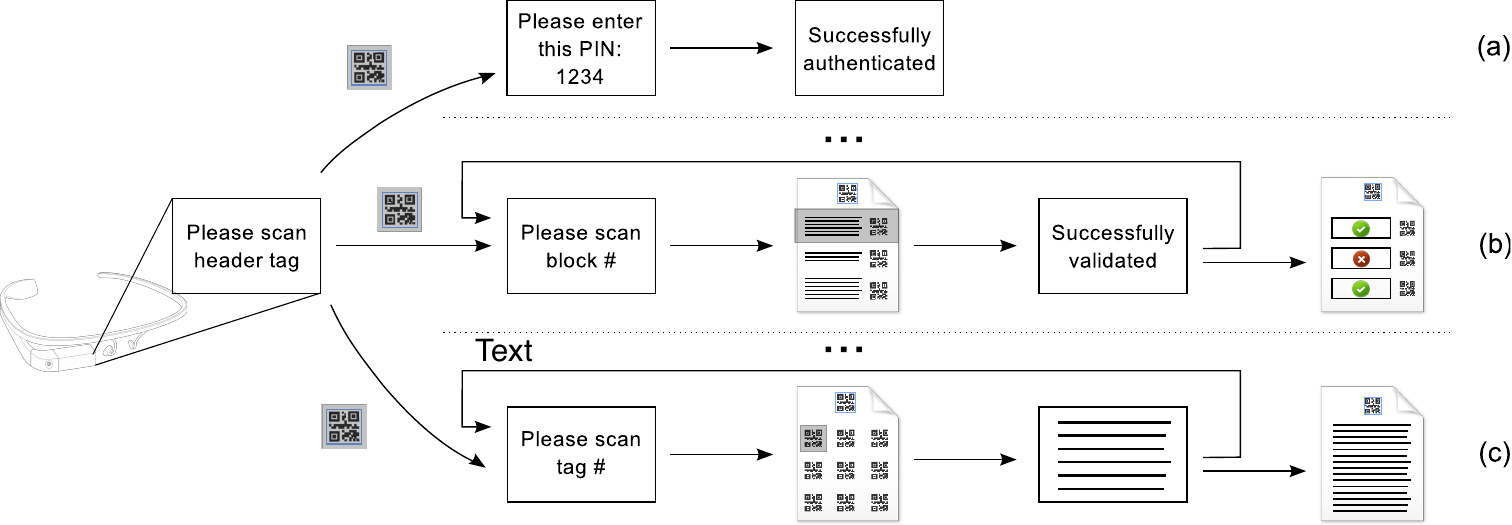}
  \caption{
Overview of the \sevi~processing and interaction pipeline for the different operation modes: identification (a), content verification (b), and content hiding (c). 
The user starts each interaction by scanning the header 2D-barcode (indicated in gray). 
\sevi~then guides the user through each process by providing constant visual feedback.
}\label{fig:overall_process}
\end{figure*}

The general processing and interaction pipeline of \sevi~is shown in Figure~\ref{fig:overall_process}. Each interaction is initialized by the user scanning the header 2D-barcode (indicated in gray). The header code is composed of the framework header and an application specific header. 
The former contains the framework version as well as the mode of operation, e.g. identification, content verification, content hiding; the latter is an application specific header, containing information that is relevant for the given application. 


\subsubsection{Assumptions}
The general setting we consider is a user who communicates with a possibly corrupted physical token over an insecure physical channel.
In this work, we concentrate on the visual channel in connection with HMDs, such as Google Glass. 
However, our framework can be adapted and extended easily to support other physical channels, such as the auditory channel, if needed.
The visual channel is very powerful and key to the vast majority of interactions that humans perform in the real-world. HMDs are personal companions that, in contrast to smartphones, sit right in front of the user's eyes. 
Google Glass comprises an egocentric camera that allows us to record the visual scene in front of the user, as well as a display mounted in front of the user's right eye. 
While the developer version that we used could still allow an observer to infer information about the content shown on the display by looking at it from the front, we assume that this is not possible in our attack scenarios. 
We consider this to be a design flaw of some of the first prototypes, which can be solved easily. 
Since the display only occludes a small corner of the user's field of view, it could simply be made opaque.
We further assume that HMDs are computationally as powerful as smartphones. 
In practice, this can be achieved by establishing a secure communication channel between the HMD and the user's smartphone. 

\renewcommand{\arraystretch}{1.3}
\begin{table}[t]
\centering
\begin{tabular}{lcccc}
\hline
\textbf{EC level} & L & M & Q & H \\
\hline
\textbf{Max. damage (\%)} & 7 & 15 & 25 & 30 \\ 
\hline
\textbf{Max. characters} & 4296 & 3391 & 2420 & 1852 \\
\hline
\end{tabular}
\caption{Maximum storage capacity for alphanumeric characters of a version 40 QR code in comparison to the error correction level and the maximum damage it can sustain.}
\label{tb:qrcodes}
\vspace{-2em}
\end{table}

An \emph{encoder} $\mathsf{E} = (\encode, \decode)$ is used to transform digital data from and to a physical representation.
We will not mention error-correcting codes explicitly, since we assume them to be a part of the encoder.
In particular, our framework uses two-dimensional barcodes, called QR codes~\cite{qrcodes}.
These codes are tailored for machine readability and use Reed-Solomon error correction~\cite{rscodes}.
Depending on the chosen error correction level, the barcode's capacity differs.
Table ~\ref{tb:qrcodes} provides a comparison of their storage capacity for alphanumeric characters and their robustness.

\section{Authentication}\label{sec:identification}
Our goal was to design an authentication mechanism that allows a user to authenticate himself in front of a token, such as a locked door or an ATM, without revealing his secret credentials to any bystanders who observe the whole authentication process.
In addition, even a malicious token should not be able to learn the user's secret credentials.
We focused on providing a solution, which is easy to deploy into the current infrastructures, i.e.\ merely a software update is required, and is as simple and user-friendly as possible.

\subsection{Threat Model}\label{sec:identification:threat}
We consider two different types of adversaries for the authentication scenario.
An \emph{active} adversary is able to actively communicate with the user and impersonate the token.
He has access to all secrets of the token itself.
His aim is to learn a sufficient amount of information about the user's credentials to impersonate him at a later point in time.
Note that security against active adversaries implies security against \emph{passive} adversaries, who are only able to observe the data that the user passes to the token during the authentication process.
Passive adversaries represent the most common real world adversaries, who can mount attacks like shoulder surfing and skimming.
A \emph{man-in-the-middle} adversary is able to misrepresent himself as the token. He is able to communicate with the user and a different token and forward possibly altered messages between the two parties.
He does not have the token's secret keys.
His aim is to authenticate in front of a different token, while communicating with the user.

\paragraph{Insecurity of current approaches}
Clearly, the most common widely deployed solutions, such as those used at ATMs, do not provide sufficient protection against such adversaries. 
During an authentication process the user's fixed PIN and card information is simply leaked to the adversary, who can then impersonate the user.

\subsection{Our Scheme}\label{sec:scheme}
\begin{figure}[t!]
\begin{minipage}{.5\textwidth}
  \includegraphics[width=\textwidth]{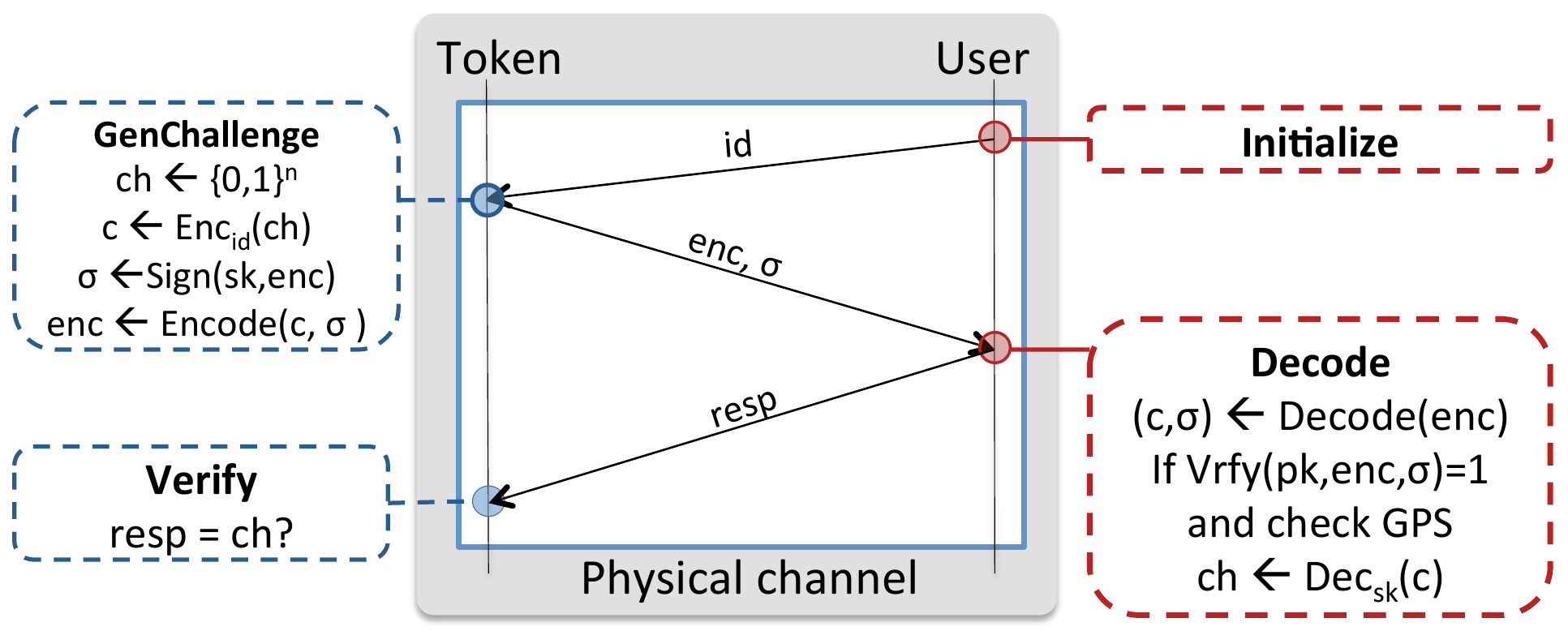}
  \caption{Visualization of a identification scheme using an optical input device.}\label{fig:identification}
\end{minipage}
\hfill
\begin{minipage}{.4\textwidth}
\centering
\begin{tabular}{|c|c|c|}
\hline
\rowcolor{lightgray}
\multicolumn{3}{|c|}{{\textsf{framework header}}}\\\hline
  \cid & \aid &\textsf{GPS} \\\hline 
\multicolumn{3}{|c|}{\encCh}\\\hline
\multicolumn{3}{|c|}{\timestamp}\\\hline
\multicolumn{3}{|c|}{\signature}\\\hline
\end{tabular}
\caption{The identification header composed of the framework and application header.}
	\label{fig:identification:header}
\end{minipage}
\vspace{-1em}
\end{figure}

Let $\pke=(\pkegen,\encrypt,\decrypt)$ be a CCA2 secure public-key encryption~\cite{katzLindell} and $\sigscheme=(\skeygen,\ssign,\sverify)$ a digital signature scheme secure against existential forgery under an adaptive chosen message attack (EU-CMA)~\cite{katzLindell}, where $\skeygen$ is the key generation algorithm, $\ssign$ is the signing algorithm, and $\sverify$, the verification algorithm.
We assume that the token has knowledge of the user's public key. In the case of an ATM, the key could be given to the bank during registration.
Our protocol is a challenge-and-response protocol that we explain with the help 
of~\autoref{fig:identification}. 
The entire communication between the user and the token uses a visual encoder, which transforms digital information to and from a visual representation.
The user initiates the protocol by sending his identifier $id$ to the token. 
The challenger retrieves the corresponding public key from a trusted database, checks the validity of the key, and encrypts a randomly generated challenge $\sf{ch}\gets\bin^n$ using the public-key encryption scheme $\pke$.
The application header for the identification scenario can be seen in~\autoref{fig:identification:header}.
It contains a token identifier (tid), a user identifier (uid), the encrypted challenge, a timestamp, and the token's GPS location.
The application header is signed with $\sigscheme$ by the token and the signature is appended to the application header.
It then generates a QR code consisting of the framework, and the application header. 

The resulting QR code is displayed to the user, who decodes the visual representation with his HMD, parses the header information, checks the validity of the signature, the date of the timestamp, whether his location matches the given location, and decrypts the encrypted challenge to obtain $\sf{ch}$.
The user sends back $\sf{ch}$ to the token to conclude the authentication process.
In the case of an ATM or a locked door, the last step can be done via a key pad.
Choosing the length of the challenge is a trade-off between security and usability.
\subsubsection{Security Analysis}\label{sec:identification:analysis}
Due to page constraints, we only provide an informal reasoning, showing that none of our three adversaries can be successful.
Note that security against the active adversary already implies security against a passive adversary.
Since we assumed that $\pke=(\pkegen, \encrypt, \decrypt)$ is secure against chosen-ciphertext attacks, an adversary is not able to infer any information about the plaintext, i.e.\ the encrypted PIN, from the given ciphertext, even if he is able to obtain encryptions and decryptions for messages of his choice.
This ensures that an (active) adversary can only guess the challenge, since he effectively plays the CCA2 game.
To prevent man-in-the-middle attacks, we use an idea called authenticated GPS coordinates, recently introduced by Marforio, Karapanos, and Soriente~\cite{2014_ndss_marforio}.
We assume that the man-in-the-middle attack is perfomed on two tokens that are at different locations. 
Recall that each token signs its challenges along with its own GPS location. 
An adversary is not able to simply forward these challenges between two tokens, since the user, upon receiving a challenge, verifies the signature of the challenge and compares its own location to the signed location.
Hence, such an adversary would need to break the unforgeability of $\sigscheme$ to be able to forward challenges that will be accepted by the user.
\section{Content Verification}\label{sec:verification}
\begin{wrapfigure}{r}{0.4\textwidth}
  \centering
  \vspace{-3em}
  \includegraphics[width=0.3\columnwidth]{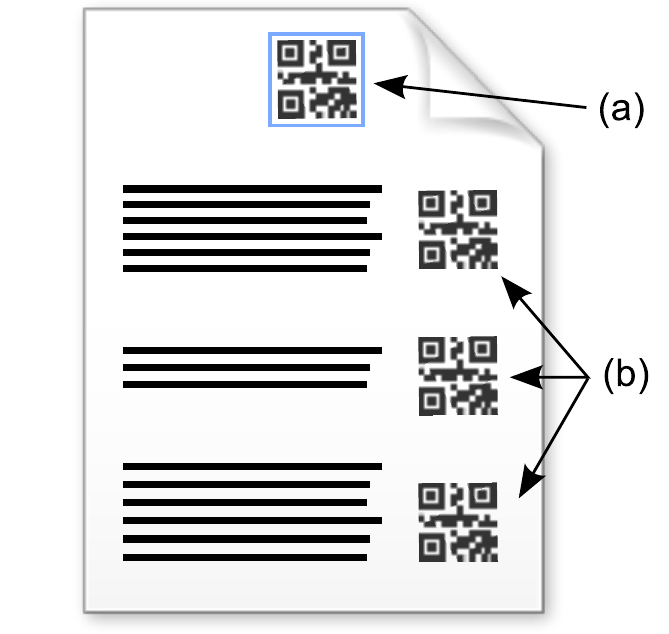}
  \caption{The \textit{\veriDoc} document format.}
  \label{fig:mach:read}
  \vspace{-0.6cm}
\end{wrapfigure}
The goal of our content verification functionality is to enable the generation and verification of physical documents, such as receipts or paychecks, with mathematically proven unforgeability guarantees.
In particular, the validity of such documents should be verifiable in a secure, user-friendly, and robust fashion.
The combination of physical documents with digital signatures is a challenging task for several reasons.
Firstly, the document's content must be human-readable, which prevents us from using machine-readable visual encodings like QR codes. Secondly, we must be able to transform the 
human readable content into a digital representation such that we can verify the digital signature. 
Here, we apply techniques from computer vision such as optical character recognition (OCR). However, OCR has 
to be performed without any errors and from a practical point of view OCR is very unlikely to succeed without any errors 
when reading a whole document with an unknown layout. Observe that error-correction techniques cannot be applied, 
since a contract that says \emph{``Alice gets \$100''} is very different from one that says \emph{``Alice 
gets \$1.00''}. 
Using error-correction one could transform a wrong document into a correct one, which would result in a discrepancy 
between what the user sees and what is verified. 
To overcome the aforementioned problems and provide a practical and useable solution, we developed a novel document format, called \veriDoc~ 
(see~\autoref{fig:mach:read}). 
This document facilitates robust document tracking and optical character recognition by encoding additional layout information into it. The
 layout information is encoded in a header QR code (a) and signatures for each block are encoded into separate QR codes (b).


\begin{figure}[t]
\begin{minipage}{.48\textwidth}
  \includegraphics[width=\textwidth]{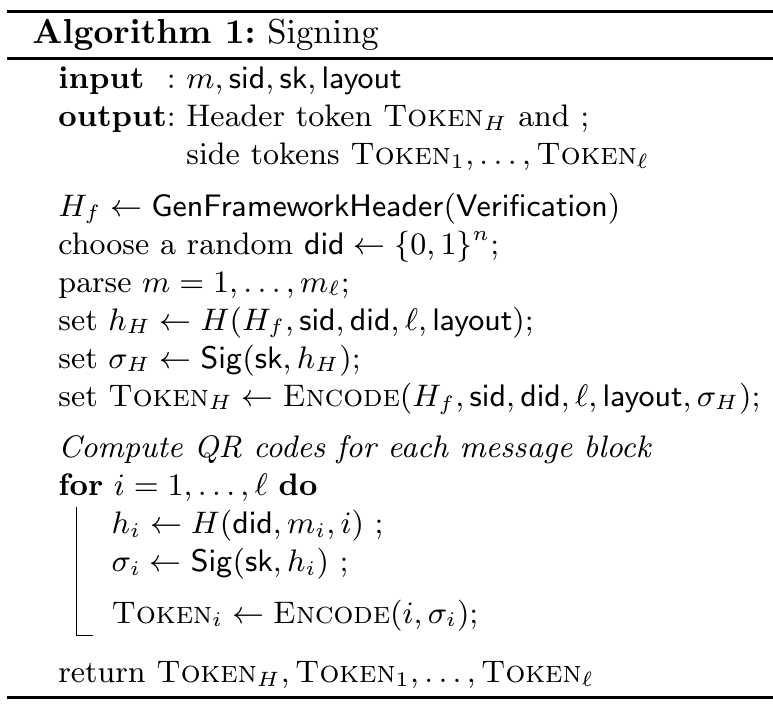}
  \caption{The signing algorithm.}\label{alg:sign}
\end{minipage}
\hfill
\begin{minipage}{.48\textwidth}
  \includegraphics[width=\linewidth]{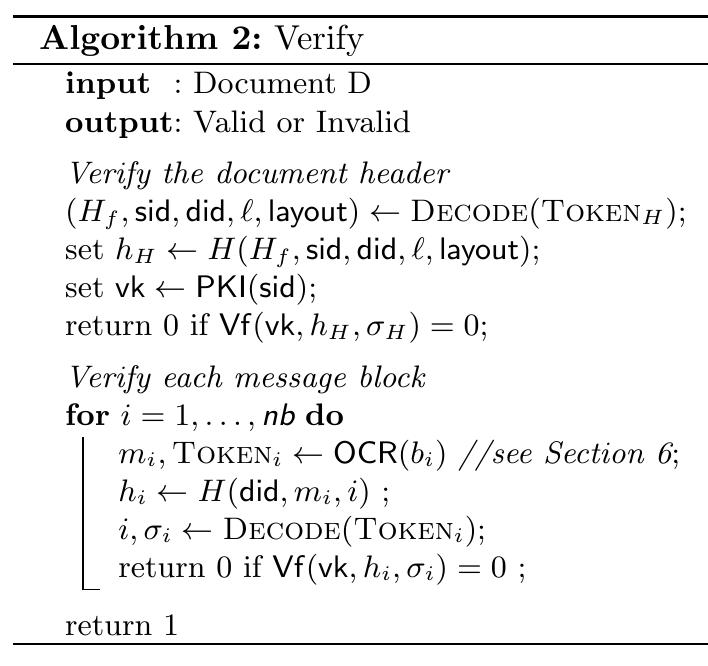}
    \caption{The verification algorithm.}\label{alg:vrfy}
\end{minipage}
  \vspace{-1em}
\end{figure}


\subsection{Threat Model}\label{sec:contvf:threat}

Based on the standard EU-CMA notion for digital signature schemes, we consider the following adversary:
In the first phase, the query phase, the adversary is able to obtain a polynomial number of (signed) \veriDoc~documents for documents of his choice from some user Alice.
In the second phase, the challenge phase, the adversary outputs a \veriDoc~document $D$ and wins if $D$ verifies under Alice's public key and was not signed by her in the first phase. 

\subsection{Our scheme}
Let $\sigscheme=(\skeygen,\allowbreak \ssign,\allowbreak\sverify)$ be a signature scheme secure against EU-CMA and $H$ a collision-resistant hash function.

\mypar{Content Signing}\label{sec:doc:encode}
A formal description of the signing algorithm is depicted in~\autoref{alg:sign}.
It takes the signer's private key $\sk$, his identifier $\sid$, the message $m=m_1,\dots,m_\ell$ consisting of $\ell$ blocks as input and the layout information
$\layout$.
First, the algorithm computes the document header $\token_H$, which comprises of the framework header and the application header.
The application header contains the signer's id $\sid$, a randomly generated document id $\did$, the number of message blocks $\ell$, and $\layout$. 
This header is signed and the signature $\sigma_H$ is appended to the header itself.
In the second step, each message block $m_i$ is signed along with $\did$ and it's position $i$.
All generated signatures are encoded into QR codes and printed onto the document next to the corresponding message blocks (see ~\autoref{fig:mach:read}).
%
%
%

\mypar{Content Verification}\label{sec:doc:decode}
The content verification algorithm, depicted in~\autoref{alg:vrfy}, is given a signed document $D$ consisting of blocks $b_i$ and verifies its validity.
The extraction of a message block, the corresponding signature, as well as the underlying computer vision techniques that are used are simplified to $\ocr(\cdot)$ in this description.
A description of $\ocr(\cdot)$ will be provided in~\autoref{sec:impl:secdoc}.
In the first step, the document header $\token_H$ is parsed by the computer vision system.
Using the signer's id $\sid$, the corresponding public key $\vk$ is obtained from a PKI.
Afterwards, the verification algorithm checks the validity of each block. To do so, the algorithm first reads the message block along with its signature $(m_i, \token_i) \exec\ocr(b_i)$, it computes the hash value $h_i \exec H(\did,m_i,i)$, extracts the signature from the corresponding QR code, i.e.\ $(i, \sigma_i)\exec\decode(\token_i)$ and outputs $0$ if the signature is invalid, i.e., if $\sverify(\vk,h_i,\sigma_i)=1$. 
If all checks are valid, then the verification algorithm outputs 1.

%
%

\subsubsection{Security Analysis}\label{sec:verification:analysis}
We assume that the underlying signature scheme $\sigscheme = (\skeygen, \ssign, \sverify)$ that is used to generate the \veriDoc~documents
is secure against EU-CMA. This means that an adversary is allowed
to obtain signatures on messages of his choice adaptively and he is not able to generate a valid signature for a new message 
that was not queried to the signing oracle before (except with negligible probability). Furthermore, we assume that the hash function is collision-resistant, meaning 
that an efficient adversary finds two distinct messages $m_0,m_1$ that map to the same image $H(m_0)=H(m_1)$ only with negligible probability. 
In the query phase, the adversary obtains signed tokens for messages of his choice. 
Note that for each signed token a new random document id $\did\in\bit^n$ is generated and the header also contains the number of blocks $\ell$. 
This document id prevents so called mix-and-match attacks, where a new valid document is generated by mixing message blocks from other valid documents. 
Since the id is $n$-bit long, where $n$ is the security parameter and we consider poly-time adversaries, the probability of two documents having the same id is negligible in $n$.
Since the signed document header contains the number of message blocks and all blocks are enumerated according to their ordering in the layout, an adversary can neither rearrange, nor remove any message blocks without breaking the unforgeability of the signature scheme. 
Thus, the resulting \veriDoc~document is also existentially unforgeable under chosen message attacks.
\subsection{Two-Step Verification}\label{sec:two_factor_verification}

Over the past years a constant increase in digital crime, such as identity theft, has been observed. 
To counteract these developments, companies like Facebook, Google, Yahoo, and many others allow users to use a technique known as two-factor authentication~\cite{TwoFacAuth}, when using their services.
During such an authentication process, an additional layer of security is introduced by requiring a second authentication factor, e.g.\ a physical token, along with the password.
In a similar vein we introduce the \emph{two-step verification technique} that introduces a second step into the process of verifying retrieved content.
Consider, for example, a user, who requests his account balance during an online banking session. 
If the machine that is used is untrusted and possibly even compromised, then the user cannot verify the correctness of the returned balance.
To overcome this problem, we use our content verification technique described in~\autoref{sec:verification}, meaning that in our banking example the account balance is returned together with a visually encoded signature thereof.
Using the HMD we parse the signature and the account balance and verify its correctness.
An adversary, who wants to convince a user of a false statement, would need to compromise the machine, that is used by him, and the HMD simultaneously, which is considerably harder to achieve in practice.
Due to the simplicity of the two-step verification technique, it could easily be integrated into many existing systems immediately.
\section{Content Hiding}\label{sec:hiding}

Motivated by the increasing existence of mobile workplaces, we introduce our content hiding solution.
Our goal was to allow users to read confidential documents in the presence of eavesdroppers.
HMDs are situated right in front of the user's eye and only he is able to see the displayed content.
Confidential documents are printed in an encrypted format and using the HMD an authorized user decrypts the part he is looking at on-the-fly.
Applications using this technique are not limited to paper-based documents or tablet computers. Consider an untrusted machine through which a user might want to access some confidential data. 
Using our content hiding technique, he could obtain the information without leaking it to the untrusted machine.
For the sake of clarity and brevity, we describe our technique using public key encryption schemes.
In~\autoref{app:access-structures} we show how to realize more complex access structures, such as security clearance hierarchies in office spaces, using predicate encryption schemes.

\subsection{Threat Model}\label{sec:hiding:threat}
In this scenario we basically consider the adversary from the standard CCA2 security notion.
The adversary is allowed to obtain a polynomial amount of encryptions and decryptions for messages and ciphertexts of his choice from some honest user Alice.
At some point the adversary outputs two messages, Alice picks one at random, and encrypts it.
The adversary wins if he can guess which message was encrypted with a probability of at least $\frac{1}{2} + \epsilon(n)$, where $\epsilon$ is a non-negligible function and $n$ is the security parameter.

\subsection{Our Scheme}\label{sec:hiding:scheme}

\begin{figure}[t]
\begin{minipage}{.48\textwidth}
  \includegraphics[width=\linewidth]{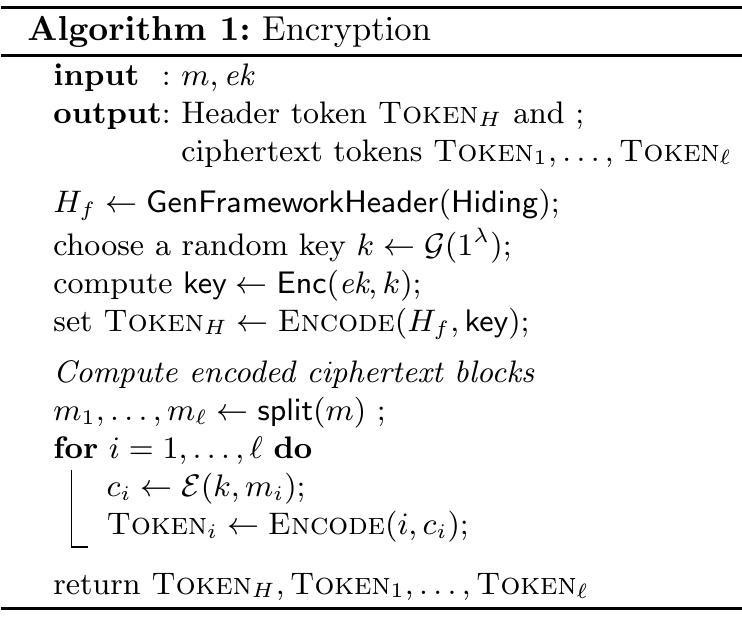}
  \caption{The encryption algorithm.}\label{alg:encrypt}
\end{minipage}
\hfill
\begin{minipage}{.48\textwidth}
  \includegraphics[width=\linewidth]{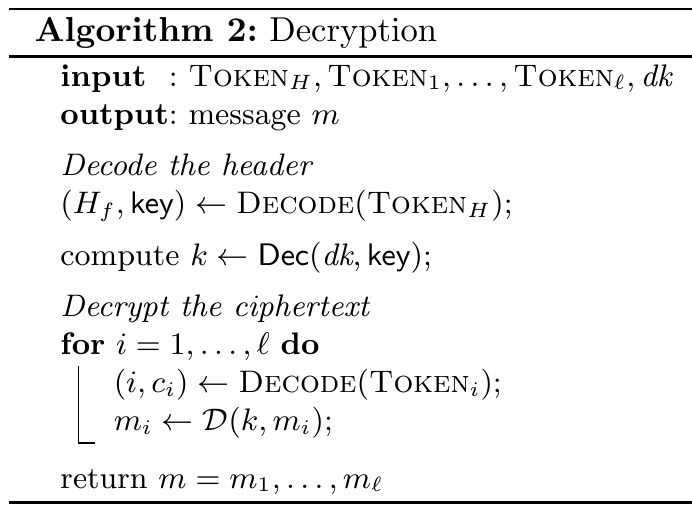}
  \vspace{0.5em}
    \caption{The decryption algorithm.}\label{alg:decrypt}
\end{minipage}
\vspace{-1em}
\end{figure}

Let $\pke=(\pkegen, \encrypt, \decrypt)$ be a CCA2 secure public key, and $\privke=(\privkegen, \privencrypt, \privdecrypt)$ a CCA2 secure private key encryption scheme.
To obtain public key encryption scheme with short ciphertexts, we use a hybrid encryption scheme~\cite{Hybrid}.
The basic idea of such a scheme is to encrypt a randomly generated key $k \exec \privkegen(n)$ with the public key encryption scheme and store it in the header.
The actual plaintext is encrypted using $\privke$ with $k$.


\mypar{Encryption} 
%
The encryption algorithm is depicted in \autoref{alg:encrypt} and works as follows:
At first, a randomly chosen \emph{document key} $k$ is encrypted with a public-key encryption scheme under the public key $\ek$ of the recipient.
A header QR code $\token_H$ is created, which contains the framework header, the encrypted document key.
The actual body of the document $m$ is split into message chunks $m_1,\dots,m_\ell$ and each chunk is encrypted separately using the document key and is then encoded, along with the block id, into a QR code $\token_i$.

\mypar{Decryption} 
The decryption algorithm is depicted in~\autoref{alg:decrypt}.
Upon receiving a document, the receiver decodes the header QR code, obtains the encrypted document key $\encKey$.
Using his secret key $\dk$, the algorithm recovers the document key $k \exec \privdecrypt(\dk, \encKey)$ and it uses the key to decrypt the document body.
%
%
%
%
%
%
%

The advantages of representing the document as a sequence of encrypted blocks is twofold.
Firstly, it allows the user to only decrypt the part of the encrypted document body that he is currently looking at without the need to scan the whole document first.
Furthermore, the encrypted documents are robust to damage, meaning that even if a part of it is broken or unreadable, we are still able to decrypt the remaining undamaged ciphertext blocks as long as the document header is readable.
Choosing the size of the message blocks is a trade-off between space and robustness. 
The bigger the message blocks are, the more plaintext is lost once a single QR code is not readable anymore. 
The smaller they are, the more QR codes are required, hence the more space is needed to display them.

\subsubsection{Security Analysis}\label{sec:hiding:analysis}
It is well known that using the hybrid argument proof technique~\cite{katzLindell} the CCA2 game, where the adversary outputs two distinct messages in the challenge phase, is equivalent to a CCA2 game where the adversary outputs two message vectors of polynomial length.
The security of our scheme directly follows from this observation. 

\subsection{Extending Content Hiding to Support Fine-Grained Access Control}\label{app:access-structures}
Using public-key encryption in our content hiding scheme allows us to encrypt documents for certain recipients. 
In companies or organizations, however, it is more desirable to encrypt documents, such that only employees with certain security clearances can read certain enrypted documents.
\sevi~allows to encrypt documents, such that only users with certain security clearances can read them.
Therefore, we replace the public-key encryption scheme by a \emph{predicate} encryption scheme~\cite{EC:KatSahWat08}. 
Loosely speaking, in a predicate encryption scheme, one can encrypt a message $M$ under a certain attribute $I \in \Sigma$ using a master public key $\mpk$ where $\Sigma$ is the universe of all possible attributes. 
The encryption algorithm outputs a ciphertext that can be decrypted with a secret key $\fmsk{f}$ associated with a predicate $f \in \mc{F}$,
if and only if $I$ fulfills $f$, i.e., $f(I) = 1$, where \mc{F} is the universe of all predicates.

Next, we explain the security notion of predicate encryption, called \emph{attribute-hiding}, with the following toy example. 
Consider the scenario where professors, students, and employees are working at a university and by \Prof, \Employee, and
\Stud we denote the corresponding attributes. 
Every member of a group will be equipped with a secret key $\fmsk{f}$ such that $f$ is either the predicate \isProf,
\isEmployee, or \isStud. We use the toy policy that professors may
read everything and employees and students may only read encryptions
created using \Employee and \Stud, respectively. 
Now, attribute-hiding states that a file $\file$ which is encrypted using the attribute
\Prof, can not be decrypted by a student equipped with $\fmsk{\isStud}$ and the 
student also can not tell with which attribute $\file$ was encrypted (except for the fact that it was not \Stud). 
Furthermore, even a professor does not learn under which attribute \file was encrypted, she only
learns the content of the file and nothing more.

\paragraph{Extending Our Scheme}
We extend our scheme to also support fine grained access control by replacing the public-key encryption scheme with a predicate encryption scheme. 
Thus, the user encrypting the message in addition chooses an attribute $I \in \Sigma$ that specifies which users can decrypt the message. 
Formally, our encryption algorithm is almost the same as described in~\autoref{alg:encrypt}, but the public-key encryption step is replaced with $c\exec\PEnc(\mpk,I,k)$, where $\mpk$ is a master public key that works for all attributes. 
The only difference in the decryption algorithm is that instead of using the public-key decryption algorithm $\decrypt$, we are 
now running the decryption algorithm of the predicate encryption scheme $\PDec(\fmsk{f},c)$ and the user
can only decrypt if $f(I) = 1$.

\paragraph{Efficient Implementation}
Our implementation is based on the predicate encryption scheme due to
Katz, Sahai, and Waters~\cite{EC:KatSahWat08} (see \autoref{sec:PredEnc} 
for a formal description of the scheme). However, for efficiency reasons, 
we did not implement the scheme in composite
order groups, but adapted the transformation to prime
order groups as suggested by Freeman~\cite{EC:Freeman10}.


\begin{figure*}[t]
\centering
\includegraphics[width=0.9\linewidth]{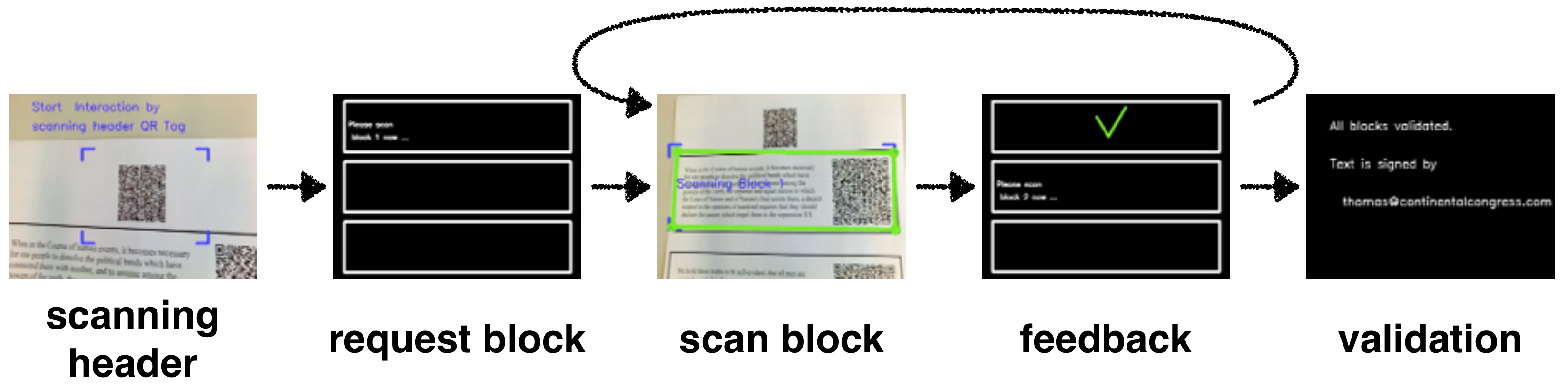}
\caption{
Interaction cycle with \veriDoc.
The user initiates the interaction by scanning the header QR code at the top of the document.
After sequential scanning of each content block, the user is informed if the document was verified or not.
The black screens are what the user sees on the Google Glass display. They cover the whole screen but only a small part of the users view.
}
\label{fig:interface}
  \vspace{-1em}
\end{figure*}

\section{The \veriDoc~Interface}\label{sec:impl:secdoc}

In the following, we describe our document format \veriDoc.
A high-level overview of the document scanning process is shown in~\autoref{fig:interface}.
Throughout this process we provide visual feedback to make the scanning process transparent to the user.
As already described, the user initiates the document verification by scanning the header code of the document.
Amongs other information, the header code contains the layout information.
This information contains additional information about the document that facilitates the scanning process.
In particular, this information contains the used font, the aspect ratio of each message block, and the document language.
After scanning the header code, the user is asked to scan the message blocks.
We display brackets on the HMD to help the user to position the camera properly over the text block
Accurate alignment and content extraction is further facilitated by a computer vision subsystem as described below. 
After each scanned block, its content is extracted and verified against the signature encoded in the QR code.
The user is informed about the validity of each text block and once all blocks of a given document are scanned the system informs the user if the document, as a whole, was successfully verified.

\begin{figure*}[t]
\centering
\includegraphics[width=0.9\linewidth]{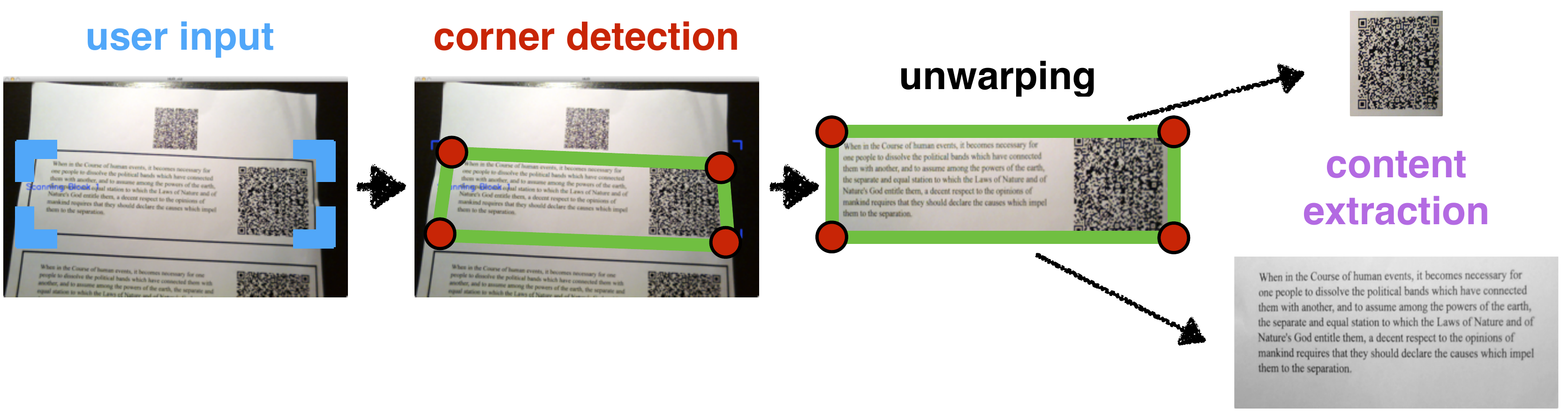}
\caption{Vision subsystem to assist the user in working with \veriDoc: The user points the front-facing camera roughly at the \veriDoc~document, the system detects the four corners of the first content block and snaps the locations of the brackets to them, and the system unwarps and extracts the content of that block.}\label{fig:vision}
  \vspace{-1em}
\end{figure*}

\mypar{Alignment and Content Extraction}
To assist the user in scanning \veriDoc~document content we provide a refinement procedure that allows the user to roughly indicate relevant text blocks, but still provide the required accuracy for the computer vision processing pipeline (see Figure \ref{fig:vision}). On the very left, a typical user interaction is depicted showing a coarse alignment of the brackets with the first text block. We proceed by a corner detection algorithm and snap the locations of the brackets provided by the user to the closest corners. We use the Harris corner measure M to robustly detect corners \cite{harris}:
\begin{eqnarray}
A &=& g(\sigma_I) * \left( 
\begin{array}{cc}
I_x^2(\sigma_D) & I_x I_y (\sigma_D) \\
I_x I_y (\sigma_D) & I_y^2(\sigma_D) \\
\end{array}
\right)\\
M &=& \det(A) - \kappa \;\mbox{trace}(A)^2
\end{eqnarray}
where $I_x$ and $I_y$ are the spatial image derivatives in x and y direction, $\sigma_D$ smoothing of the image with the detection scale  and $\sigma_I$ smoothing the response with the integration scale and $\kappa = 0.04$ according to best practice. Intuitively, the pre-smoothing with $\sigma_D$ eliminates noise and allows detection of corners at a desired scale \cite{lindeberg} while the smoothing $\sigma_I$ suppresses local maxima in the response function. In order to be robust to the choice of these scales, we employ the multi-scale harris detector that finds corners across multiple scales \cite{mikolajczyk05}.

The second image at the top shows a visualization of the closest corner and the box spanned by them in green. 
Under the assumption of a pinhole camera model as well as a planar target (documents in our case), we can compute a homography $H \in \mathbb{R}^{3\times 3}$ in order to undo the perspective transformation under which the content is viewed. The matrix $H$ relates the points under the perspective project $p'$ to the points under an orthogonal viewing angle $p$ by
\begin{eqnarray}
p' = H p
\end{eqnarray}
where $p,p' \in \mathbb{R}^3$ are given in homogeneous coordinates.
As our interface has determined the $4$ corners that each specify a pair of $p$ and $p'$, we have sufficient information to estimate matrix $H$.

The third image from the left in Figure~\ref{fig:vision} shows the content after unwarping and cropping. Using the information on the ratio between text and code contained in the header, we now split the content area into text and the associated QR code. 
In a last step, the QR code is decoded, the signature extracted, and the text area is further processed using OCR in combination with the font and language information from the header code.

\section{Prototype Implementation}\label{sec:implementation}

We provide a prototype implementation, written in Java, of our \sevi~framework on the Google Glass device.
The device runs Android 4.0.4 as its underlying operating system, features a 640$\times$360 optical head-mounted display as well as an egocentric camera with a resolution of 1280$\times$720 pixels. 
Our current developer version only features an embedded microcontroller with 1.2 GHz and 1GB of memory. 


We used the Bouncy Castle Crypto API 1.50~\cite{bouncy} and the Java Pairing-Based Cryptography Library 2.0.0 (JPBC)~\cite{ISCC:DecIov11} to implement all required cryptographic primitives.
In particular, we used SHA-1 as our collision-resistant hash function, SHA1+RSA-PSS as our signature, AES-256 in CTR-Mode as our private-key encryption, and RSA-OAEP with 2048 bit long keys as our public-key encryption scheme.
For our predicate encryption scheme, we use a MNT curve~\cite{Miyaji01newexplicit} with a security parameter of 112 according to the NIST recommendations~\cite{nistKey}.
For the computer vision part of our framework, we used the OpenCV 2.4.8 image processing library~\cite{opencv}, and QR codes are being processed with the barcode image processing library zxing 1.7.6~\cite{zxing2012}.
For optical character recognition, we used the Tesseract OCR engine~\cite{tesseract}.


\section{Related Work}\label{sec:related_work}

Head-mounted displays, such as Google Glass, have raised strong privacy concerns in the past and recent publications~\cite{osSupportAR,AScannerDarkly,ARRecognizer} have tried to address these issues.
In~\cite{osSupportAR}, the authors suggest that operating systems should provide high-level abstractions for accessing perceptual data.
Following this line of work,~\cite{AScannerDarkly} proposes a system, which makes a first step towards providing privacy guarantees for sensor feeds.
There, applications access the camera through a new interface rather than accessing the camera directly.
Depending on the application's permissions, the camera is pre-processed with different sets of image filters, which aim to filter sensitive information.
In~\cite{ARRecognizer}, the notion of recognizers is introduced. 
Rather than passing a filtered sensor feed to the requesting application, they provide a set of recognizers that fulfil the most common tasks, such as face detection or recognition. 
Applications obtain permissions for certain recognizer and can request the output of certain computations on the sensor feed.
In contrast to our work, this line of research regards the device as a threat. 

Another line of research concentrates on establishing trust between devices based on the visual channel \cite{QRTAN,SecureDevicePairingVisualChannel,SeeingIsBelieving}.
In~\cite{SeeingIsBelieving}, for instance, the visual channel is used for demonstrative authentication, where two devices authenticate themselves towards each other by basing their trust on the visual channel between them.
One possible application for this authentication mechanism is access points with QR codes printed onto them. 
The user scans the QR code to authenticate the access point.

In~\cite{scanTextSurvey}, a survey of different techniques for scanning and analyzing documents with the help of cameras, cell phones, and wearable computers is provided.
The survey shows that even though constant progress is made, current methods are not robust enough for real world deployment.
\sevi~tackles this problem in a different way by facilitating the task of scanning documents by encoding additional information into them.

\section{Conclusion}

We presented \sevi, a framework that makes an important step towards a tighter integration of digital cryptography in real-world applications.
Using HMDs in combination with established cryptographic primitives and resource friendly computer vision techniques, we provide users with more security and privacy guarantees in a wide range of common real-world applications.
We present user-friendly, easy-to-use solutions for authentication, content verification, content hiding, that can seamlessly be integrated into the current infrastructure.
We hope that our work will stimulate further research investigating the possibilities of combining ubiquitous computing technologies with cryptographic primitives in a user-friendly fashion.

\subsection*{Acknowledgements}
Andreas Bulling and Mario Fritz are supported by a Google Glass Research Award.
Work of Dominique Schr\"oder and Mark Simkin 
was supported by the German Federal Ministry of Education and Research
(BMBF) through funding for the Center for IT-Security, Privacy, and Accountability (CISPA; see
\texttt{www.cispa-security.org}). Dominique Schr\"oder is also supported by an Intel Early Career Faculty Honor Program Award.





\bibliographystyle{IEEEtran}
\bibliography{bib/extrarefs,bib/visionhci}

\appendix
\section{Predicate Encryption} \label{sec:PredEnc}
For completeness, we recall the predicate encryption scheme due to
Katz, Sahai, and Waters~\cite{EC:KatSahWat08}. 

\begin{definition}[Predicate
  Encryption]\label{def:predicate-encryption}
  A \emph{predicate encryption scheme} for the universe of predicates
  and attributes \mc{F} and $\Sigma$, respectively, is a tuple of efficient
  algorithms $\PE = (\PGen, \PKeyGen, \PEnc, \PDec)$, where the
  generation algorithm \PGen takes as input a security parameter
  $1^\secparam$ and returns a master public and a master secret key
  pair $(\mpk, \msk)$; the key generation algorithm \PKeyGen takes as
  input the master secret key \msk and a predicate description $f \in
  \mc{F}$ and returns a secret key $\fmsk{f}$ associated with $f$; the
  encryption algorithm \PEnc takes as input the master public key
  \mpk, an attribute $I \in \Sigma$, and a message $m$ and it returns
  a ciphertext $c$; and the decryption algorithm \PDec takes as input
  a secret key $\fmsk{f}$ associated with a predicate $f$ and a
  ciphertext $c$ and outputs either a message $m$ or $\bot$.
\end{definition}

A predicate encryption scheme \PE is \emph{correct} if and only if,
for all \secparam, all key pairs $(\mpk, \msk) \gets
\PGen(1^\secparam)$, all predicates $f \in \mc{F}$, all secret keys
$\fmsk{f} \gets \PKeyGen(\msk, f)$, and all attributes $I \in \Sigma$
we have that $(i)$ if $f(I) = 1$ then $\PDec(\fmsk{f}, \PEnc(\mpk, I,
m)) = m$ and $(ii)$ if $f(I) = 0$ then $\PDec(\fmsk{f}, \PEnc(\mpk, I,
m)) = \bot$ except with negligible probability.

\paragraph{The KSW Predicate Encryption Scheme} \label{sec:PredEncDesc}
\label{sec:ksw-pred-encrypt}

The scheme is based on composite order groups with a bilinear
map. More precisely, let $N = pqr$ be a composite number where $p$,
$q$, and $r$ are large prime numbers. Let $\mbb{G}$ be an order-$N$
cyclic group and $e : \mbb{G} \times \mbb{G} \rightarrow \mbb{G}_T$ be
a bilinear map. Recall that $e$ is \emph{bilinear}, i.e., $e(g^a, g^b)
= e(g,g)^{ab}$, and \emph{non-degenerate}, i.e., if $\generator{g} =
\mbb{G}$ then $e(g,g) \neq 1$. Then, by the chinese remainder theorem,
$\mbb{G} = \mbb{G}_p \times \mbb{G}_q \times \mbb{G}_r$ where
$\mbb{G}_s$ with $s \in \set{p,q,r}$ are the $s$-order subgroups of
$\mbb{G}$. Moreover, given a generator $g$ for $\mbb{G}$,
$\generator{g^{pq}} = \mbb{G}_r$, $\generator{g^{pr}} = \mbb{G}_q$,
and $\generator{g^{qr}} = \mbb{G}_p$. Another insight is the
following, given for instance $a \in \mbb{G}_p$ and $b \in \mbb{G}_q$,
we have $e(a,b) = e((g^{qr})^c, (g^{pr})^d) = e(g^{rc}, g^d)^{pqr} =
1$, i.e., a pairing of elements from different subgroups cancels
out. Finally, let \mc{G} be an algorithm that takes as input a
security parameter $1^\secparam$ and outputs a description $(p, q, r,
\mbb{G}, \mbb{G}_T, e)$. We describe the algorithms \PGen, \PKeyGen,
\PEnc, and \PDec in the sequel.

\paragraph{Algorithm $\POGen(1^\secparam, n)$ and $\PGen(1^\secparam,
  n)$}
First, the algorithm runs $\mc{G}(1^\secparam)$ to obtain $(p,q,r,
\mbb{G}, \mbb{G}_T, e)$ with $\mbb{G} = \mbb{G}_p \times \mbb{G}_q
\times \mbb{G}_r$. Then, it computes $g_p$, $g_q$, and $g_r$ as
generators of $\mbb{G}_p$, $\mbb{G}_q$, and $\mbb{G}_r$,
respectively. The algorithm selects $R_0 \in \mbb{G}_r$, $R_{1.i},
R_{2,i} \in \mbb{G}_r$ and $h_{1,i}, h_{2,i} \in \mbb{G}_p$ uniformly
at random for $1 \le i \le n$. $(N = pqr, \mbb{G}, \mbb{G}_T, e)$
constitutes the public parameters. The public key for the
predicate-only encryption scheme is
\[
\begin{array}{l}
  \mopk = (g_p, g_r, Q = g_q \cdot R_0,\set{H_{1,i} = h_{1,i} \cdot
  R_{1,i}, H_{2,i} = h_{2,i} \cdot R_{2,i}}_{i=1}^{n})
\end{array}
\]
and the master secret key is
$
\mosk = (p,q,r, g_q, \set{h_{1,i}, h_{2,i}}_{i=1}^n).
$
For the predicate encryption with messages, the algorithm additionally
chooses $\gamma \in \mbb{Z}_N$ and $h \in \mbb{G}_p$ at random. The
public key is
\[
\begin{array}{l}
  \mpk = (g_p, g_r, Q = g_q \cdot R_0, P = e(g_p, h)^\gamma,
\set{H_{1,i} = h_{1,i} \cdot
    R_{1,i}, H_{2,i} = h_{2,i} \cdot R_{2,i}}_{i=1}^{n})
\end{array}
\]
and the master secret key is
$
\msk = (p,q,r, g_q, h^{-\gamma}, \set{h_{1,i}, h_{2,i}}_{i=1}^n).
$

\paragraph{Algorithm $\POKeyGen(\mosk, \vec{v})$ and $\PKeyGen(\msk,
  \vec{v})$} 
Parse $\vec{v}$ as $(v_1, \ldots, v_n)$ where $v_i \in \mbb{Z}_N$. The
algorithm picks random $r_{1,i}, r_{2,i} \in \mbb{Z}_p$ for $1 \le i
\le n$, random $R_5 \in \mbb{G}_r$, random $f_1, f_2 \in \mbb{Z}_q$,
and random $Q_6 \in \mbb{G}_q$. For the predicate-only encryption
scheme, it outputs a secret key
\[
\fmosk{\vec{v}} = \left(
\begin{array}{c}
  K_0 = R_5 \cdot Q_6 \cdot \prod_{i=1}^n h_{1,i}^{-r_{1,i}} \cdot
  h_{2,i}^{-r_{2,i}},\\[2mm]
  \set{K_{1,i} = g_p^{r_{1,i}} \cdot g_q^{f_1 \cdot v_i}, 
    K_{2,i} = g_p^{r_{2,i}} \cdot g_q^{f_2 \cdot v_i}}_{i=1}^n
\end{array}
\right).
\]
For the predicate encryption scheme with messages, the secret key
\fmsk{\vec{v}} is the same as \fmosk{\vec{v}} except for 
\[
K_0 = R_5 \cdot Q_6 \cdot h^{-\gamma} \cdot \prod_{i=1}^n
h_{1,i}^{-r_{1,i}} \cdot h_{2,i}^{-r_{2,i}}.
\]

\paragraph{Algorithm $\POEnc(\mopk, \vec{x})$ and $\PEnc(\mpk,
  \vec{x}, m)$}
Parse $\vec{x}$ as $(x_1, \ldots, x_n)$ where $x_i \in \mbb{Z}_N$. The
algorithm picks random $s, \alpha, \beta \in \mbb{Z}_N$ and random $R_{3,i},
R_{4,i} \in \mbb{G}_r$ for $1 \le i \le n$. For the predicate-only
encryption scheme, it outputs the ciphertext
\[
C = \left(
\begin{array}{c}
C_0 = g_p^s, \{C_{1,i} = H_{1,i}^s \cdot Q^{\alpha \cdot x_i} \cdot
R_{3,i},\\[2mm] 
C_{2,i} = H_{2,i}^s \cdot Q^{\beta \cdot x_i} \cdot R_{4,i} \}_{i=0}^n
\end{array}
\right).
\]
For the predicate encryption scheme with messages notice that $m \in
\mbb{G}_T$. The ciphertext is
\[
C = \left(
\begin{array}{c}
C' = m \cdot P^s, C_0 = g_p^s,\\[2mm]
 \{C_{1,i} = H_{1,i}^s \cdot Q^{\alpha \cdot x_i} \cdot R_{3,i},\\[2mm] 
C_{2,i} = H_{2,i}^s \cdot Q^{\beta \cdot x_i} \cdot R_{4,i} \}_{i=0}^n
\end{array}
\right).
\]

\paragraph{Algorithm $\PODec(\fmosk{\vec{v}}, C)$ and
  $\PDec(\fmsk{\vec{v}}, C)$}
The predicate-only encryption outputs whether the following equation is
equal to $1$
\[
e(C_0, K_0) \cdot \prod_{i=1}^n e(C_{1,i}, K_{1,i}) \cdot e(C_{2,i}, K_{2,i}).
\]
The predicate encryption scheme with messages outputs the result of
the following equation
\[
C' \cdot e(C_0, K_0) \cdot \prod_{i=1}^n e(C_{1,i}, K_{1,i}) \cdot
e(C_{2,i}, K_{2,i}).
\]

\end{document}